\documentclass[twocolumn]{emulateapj}
\usepackage{amsmath}
\usepackage{natbib}

\slugcomment{To appear in PASP}

\shorttitle{Characterizing Reciprocity Failure}
\shortauthors{Biesiadzinski \textit{et al.}}

\begin{document}


\title{Measurement of Reciprocity Failure in Near Infrared Detectors 
}


\author{
T.~Biesiadzinski,
W.~Lorenzon, R.~Newman,
M.~Schubnell, G.~Tarl\'e,
C.~Weaverdyck}
\affil{Department of Physics, University of Michigan,
Ann Arbor, MI 48109}


%
%
\begin{abstract}
Flux dependent non-linearity (reciprocity failure) in HgCdTe near infrared
detectors can severely impact an instrument's performance, in particular with
respect to precision photometric measurements. The cause of this effect is
presently not understood. To investigate reciprocity failure, a dedicated test
system was built. For flux levels between 1 and 50,000 photons/s, a sensitivity
to reciprocity failure of approximately 0.1\%/decade was achieved. A wavelength
independent non-linearity due to reciprocity failure of about 0.35\%/decade was
measured in a 1.7$\,\mu$m HgCdTe detector.
\end{abstract}


\keywords{cosmology -- photometry -- astronomical instrumentation}
%

\section{Introduction}
Near infrared (NIR) detector technology has made great strides over the past
two decades and large format arrays with excellent performance are now
commercially available. Substrate-removed devices extend the wavelength
sensitivity of near infrared detectors into the UV and highly integrated
read-out ASICS provide compact, low power front-end electronics. Advances in
detector technology make NIR detectors well suited for space-based wide-field
imaging instruments, which are critical for pursuing some of the major
scientific questions of our time. One of the most far-reaching  problems in
physics today is the lack of understanding of the nature of dark energy. The
investigation of dark energy is most efficiently pursued with experiments that
employ a combination of different observational probes, such as type-Ia
supernovae, weak gravitational lensing, galaxy and galaxy cluster surveys, and
baryon acoustic oscillations. Most of these approaches rely on photometric
calibrations over a wide range of intensities using standardized stars and
internal reference sources. Hence, a complete understanding of the linearity of
the detectors is necessary. As part of a comprehensive program to study HgCdTe
detector properties that impact precision photometry, we have studied flux
dependent detector non-linearity. This effect was observed in the Near
Infra-Red Camera and Multi-Object Spectrometer (NICMOS) on the Hubble Space
Telescope (HST)~\citep{NICMOS0502, NICMOSoverview}. The NICMOS instrument,
installed onboard HST during the second servicing mission in 1997, employs
three 256\,$\times$\,256 NIR detectors. These $2.5\,\mu$m cut-off HgCdTe
devices were fabricated by Rockwell Science Center (now Teledyne Imaging
Sensors, TIS). This vendor also supplied the 1024\,$\times$\,1024 $1.7\,\mu$m
cut-off HgCdTe detector for the Wide Field Camera\,3 (WFC3) instrument
\citep{Baggett}, which was recently installed on HST during the final servicing
mission. The $1.7\,\mu$m cut-off HgCdTe detector used for the reciprocity study
described here was also supplied by TIS.

The NICMOS team concluded that the NICMOS detectors exhibit a significant flux
dependent non-linearity which strongly varies with wavelength
\citep{NICMOS0502}. This non-linearity, referred to here as ``reciprocity
failure'', must be carefully distinguished from the well-known non-linearity of
total signal, referred to here as ``classical non-linearity'', which is
observed in near infrared detectors that integrate charge on the junction
capacitance of the pixels. Classical non-linearity in NIR detectors is caused
by dependence of diode capacitance on voltage and non-linearity in the readout
multiplexer, and is usually measured by integrating a constant flux for
different exposure times. Reciprocity failure in turn can be measured by
varying the flux for exposure times that produce a constant integrated signal.

%

The mechanism responsible for reciprocity failure is not yet understood. It has
been suggested that image persistence in HgCdTe detectors is caused by the slow
release of trapped charge in the bulk material (Smith et al. 2008). It is
conceivable that charge traps are also the cause of reciprocity failure. For a
trap density that is small but not negligible compared to the photon density at
low illumination levels, a small fraction of the signal would be lost due to
the traps. An increase in the photon flux then will not result in a
proportionally reduced signal, since charge is not efficiently exposed to traps
with long fill-time constants. However, longer illumination at low flux levels
could result in the filling of traps with long fill time constants and thus a
reduced integrated signal. Such a detector behavior would produce the observed
effect: for a given total integrated signal a pixel's response to a high flux
is larger than to a low flux. Mathematically, reciprocity failure can be
characterized by a logarithmic behavior over most of the dynamic range of a
detector and the deviation from a linear system is expressed as fractional
deviation per decade of total signal response.

Reciprocity failure impacts photometry as residual pixel-level uncertainties
directly propagate to the estimated uncertainty on the derived magnitude.
Detailed knowledge of the degree of reciprocity failure for a detector will
affect the calibration strategy and the calibration devices needed. A profound
understanding of the cause of this effect could influence the detector
manufacturing process, possibly reducing or even eliminating this
non-linearity.
%

\section{Instrument}
\label{instrument_section}

To quantify reciprocity failure in NIR detectors, a dedicated test system was
designed and built. Based on the measurements reported by the NICMOS team it
was determined that a sensitivity to reciprocity failure of at least 1\%/decade
over the full dynamic range of a typical NIR detector had to be achieved. To
measure reciprocity failure a detector was exposed at different illumination
intensities, and the incident flux was precisely monitored with photodiodes.
The exposure time at each illumination intensity was adjusted to integrate to
similar total integrated signals. A parametrization including classical
non-linearity and reciprocity failure was used to describe the data and to
extract a measurement of the non-linearity due to reciprocity failure (see
Appendix). Knowledge of the linearity of the photodiodes is essential to this
method. Therefore, deviation from linearity of the photodiodes was measured
independently as described in Section\,\ref{section_photodiodes}.
\begin{figure}[h]
\includegraphics[width=0.98\linewidth]{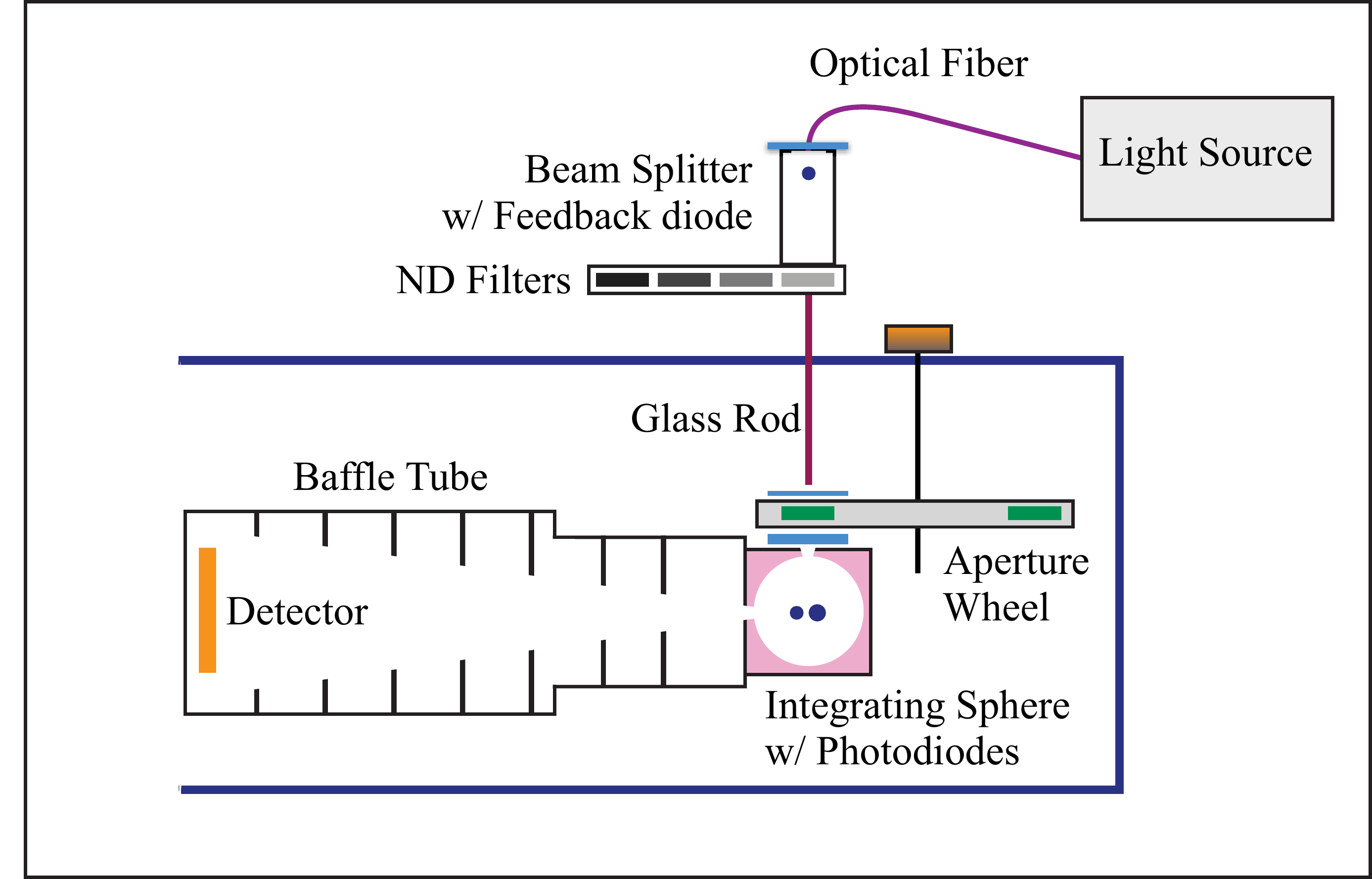}
\caption{Schematic overview of the set-up used to measure reciprocity failure.
Not shown is the liquid nitrogen vessel to which this set-up is attached.\\} \label{fig:setup}
\end{figure}

The experimental set-up utilizes a fixed illumination geometry. The
illumination intensity is varied through a combination of neutral density
filters and pinhole apertures, as schematically illustrated in
Fig.\,\ref{fig:setup}. A regulated light source placed outside the dewar is
connected via a liquid light guide to a glass rod that illuminates a pinhole
mounted on the aperture wheel inside the dewar. To avoid stray light entering
the dewar, the glass rod is surrounded by a bellows that attaches to the cold
shield and the aperture wheel. The detector is illuminated by an integrating
sphere, placed immediately below the aperture wheel, with fixed aperture and
baffling. This produces an illumination profile at the detector that is
independent of illumination intensity. The baffle tube, located between the
integrating sphere and the detector, prevents stray light and reflected light
from reaching the detector and keeps the illuminating geometry fixed. A set of
six pinhole apertures at the input of the integrating sphere combined with
neutral density filters at the entrance of the dewar extension allow a dynamic
range in intensity of approximately $10^6$ to be covered. Because all
measurements are relative to the photodiodes that monitor the incident flux,
knowledge of the exact area of the pinholes is not critical. Furthermore,
knowledge of the exact optical densities of the neutral density filters is also
not essential. Since neutral density filters can show spectral dependence,
pinhole apertures were used to verify the spectral flatness of the ND filters
utilized in the set-up at a level sufficient for the measurements reported
here.

\subsection{Illumination}

The detector inside the dewar is illuminated by one of two light sources: a
feedback controlled 50\,W Quartz-Tungsten-Halogen (QTH) lamp or alternatively a
790\,nm diode laser. Light from the QTH light source is guided by a liquid
light guide (Newport 77634) to a 70/30 beam splitter for feedback diode
pick-up. A Si feedback diode connected to the QTH lamp control electronics
stabilizes the QTH light source. Bulbs were changed frequently to avoid
end-of-life fluctuations and spectral variations. A filter stack in front of
the beam splitter provides for pass-band selection. Depending on the wavelength
selected for the measurement, either a 900\,nm long-pass filter or a stack of a
1100\,nm short-pass filter and a 1000\,nm short-pass filter (to improve
out-of-band blocking) was inserted into the light path. The pass filter is then
followed by one of four band-pass filters.\footnote{The following band-pass
filters were used: 700 nm central wavelength, 80\,nm wide; 880\,nm, 50\,nm
wide; 950\,nm, 50\,nm wide; and 1400\,nm, 80\,nm wide.} Following the splitter,
the re-focussed light beam passes through a filter slide, housing a selectable
set of neutral density filters with optical densities 0, 1, 2, and 3. The
connection from the warm optics into the dewar is made by a glass rod. Light
from the glass rod is then incident on the selected aperture inside the
aperture wheel. The aperture wheel has a total of eight positions, six of which
house pinholes ranging in diameter from 30\,$\mu$m to 11\,mm (30\,$\mu$m,
100\,$\mu$m, 330\,$\mu$m, 1\,mm, 3.3\,mm, and 11\,mm), one position completely
blocks the light, and one position is fully open with no aperture
($\approx$\,13\,mm diameter).

The pinhole illuminates the entrance port of a 2-inch integrating sphere
(SphereOptics SPH-2Z-4) as shown in Fig.\,\ref{fig:setup}. An optional
short-pass cold filter (Asahi YSZ1100) between two diffusers just in front of
the integrating sphere is used for measurements below 1000\,nm. The inside of
the integrating sphere is coated with polytetrafluoroethylene (PTFE) based
material providing good reflectivity at NIR wavelengths and good low
temperature performance.

\subsection{Photodiode Calibration}
\label{section_photodiodes}
%
The reciprocity set-up was designed for measurement of substrate removed NIR
HgCdTe detectors which exhibit spectral response at visible and NIR
wavelengths. Two photodiodes, an InGaAs photodiode and a Si photodiode, were
selected for good wavelength coverage. The NIR photodiode is a blue extended
InGaAs PIN diode (Hamamatsu Photonics G108799-01K) with an effective area of
0.785 mm$^2$ and spectral response range of 0.5\,$\mu$m to 1.7\,$\mu$m. For
improved sensitivity in the visible, a Si photodiode (Edmund Optics 53371) with
an effective area of 5.1\,mm$^2$ and spectral response between 0.5\,$\mu$m and
1.1\,$\mu$m was used. The two photodiodes were mounted adjacent to each other
to an open port of the integrating sphere as shown in Fig.\,\ref{fig:setup} and
were read out in parallel.

\begin{figure}[ht]
\includegraphics[width=0.98\linewidth]{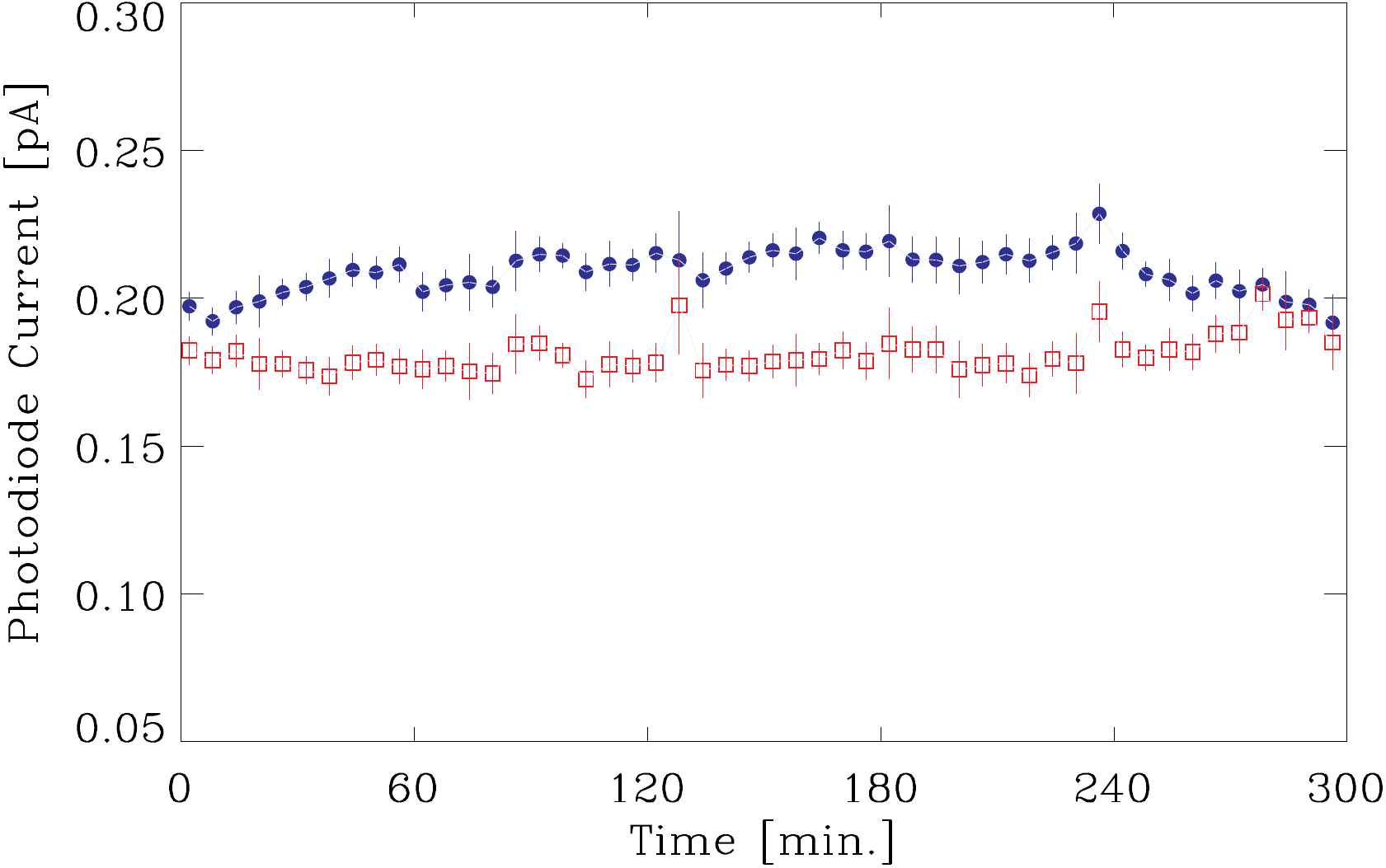}
\caption[photodiode]
{\label{fig:photodiode}
InGaAs photodiode current as a function of time. The blue circles show the
time averaged dark corrected current registered during a reciprocity
measurement extending over 5 hours. The red squares show the same photodiode
measurement corrected for fluctuations of the pico-ammeter.}
\end{figure}

The photodiode currents were recorded by two Keithley 6485 pico-ammeters that
were read out through a GPIB interface by the data acquisition computer. For
stable performance, the pico-ammeter was turned on at least 1 hour prior to
every series of measurements. Typical photodiode currents were of order 1\,pA
to 10\,nA for the InGaAs photodiode and 10\,pA to 100\,nA for the Si
photodiode. An accurate photodiode current measurement requires multiple
samples. This was achieved by operating the pico-ammeter in sampling mode and
by averaging over ten such samplings. Instrument drift during very long
exposures was tracked by a reference photodiode and subtracted from the
photodiode signal as shown in Fig.\,\ref{fig:photodiode}.

\begin{figure}[h]
\centerline{\includegraphics[width=0.9\linewidth]{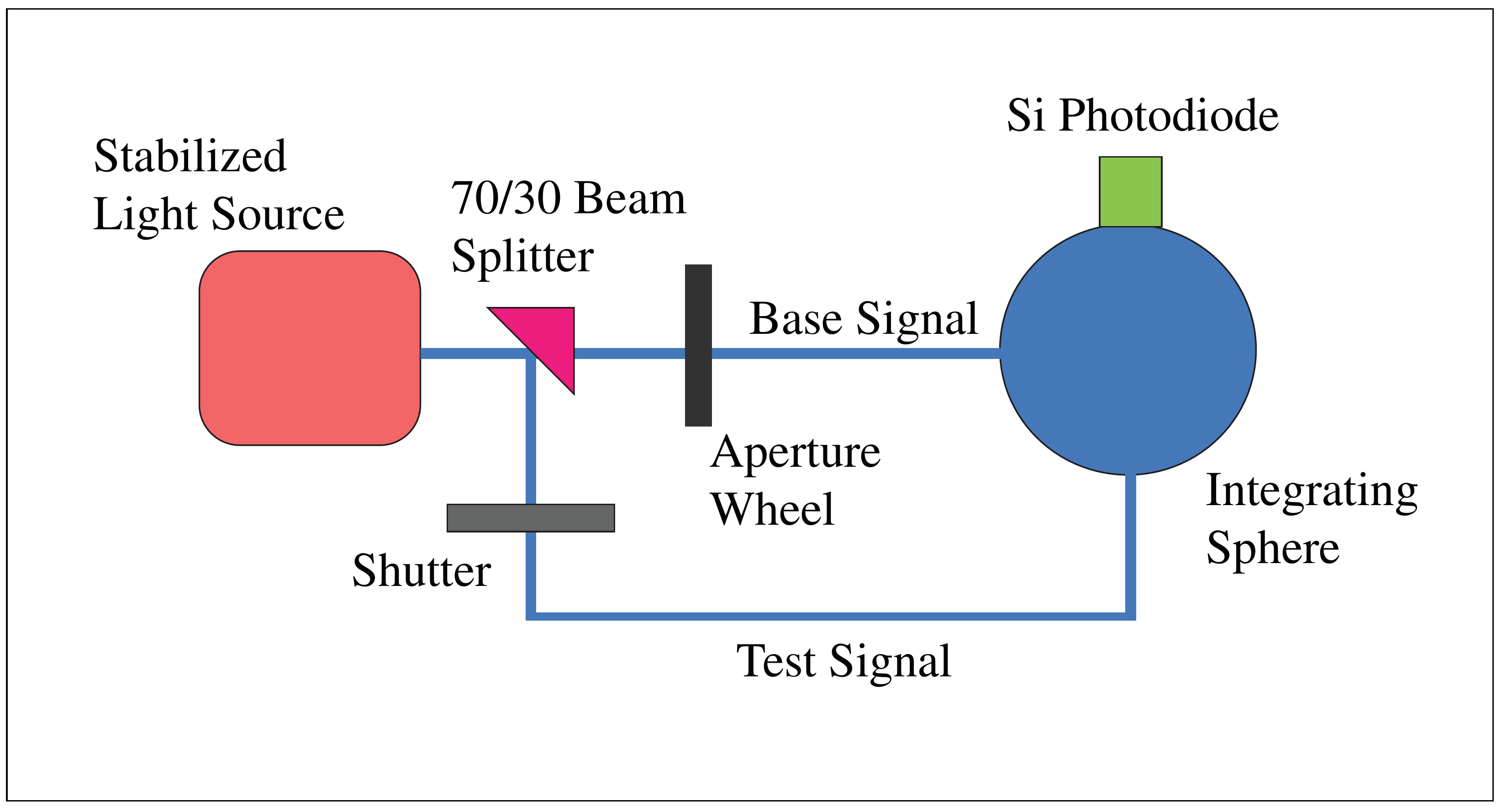}}
\caption[diode_linearity_set_up]
{\label{fig:diode_linearity_set_up}
Schematic set-up used to measure Si photodiode linearity.}
\end{figure}

\begin{figure}[h]
\includegraphics[width=0.98\linewidth]{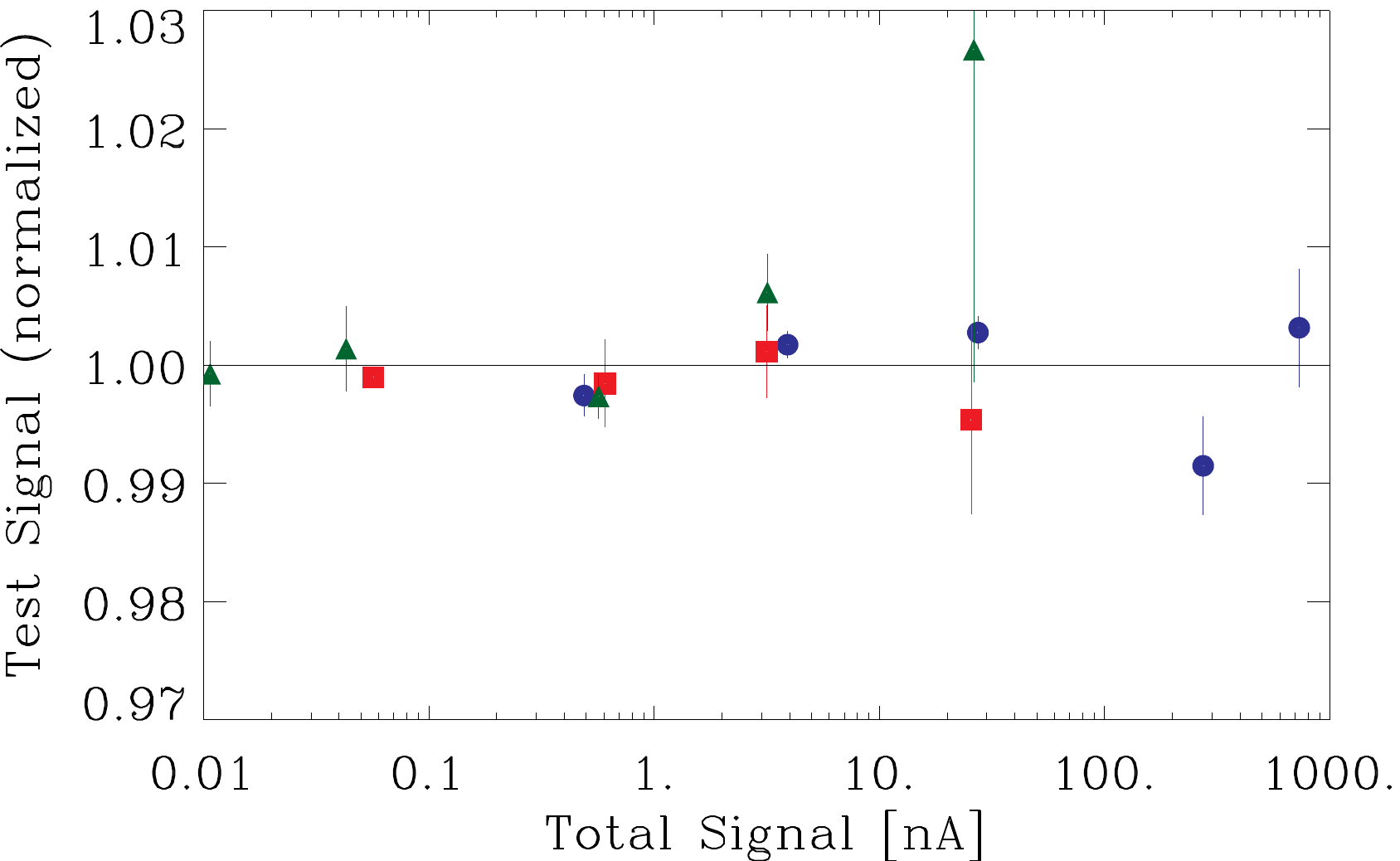}
\caption[diode_linearity]
{\label{fig:diode_linearity}
Normalized test signal as a function of total signal (base signal plus test signal) at
approximately 9\,pA (green triangles), 55\,pA (red squares), and 488\,pA (blue circles). A combined fit to all
data results in a non-linearity of (0.08\,$\pm$\,0.08)\%/decade for the Si photodiode. Note that the error
bars on the normalized test signals represent mainly the systematic uncertainties in these
measurements, since the statistical uncertainties are negligible in comparison. }
\end{figure}

Our measurement technique requires that any deviation from photodiode linearity
be well characterized and corrected for. Since precise linearity specifications
were not available from the photodiode vendors, photodiode linearity  was
measured in our laboratory. We used a beam-addition method in which a small,
constant ``test signal'' was intermittently added to ``base signals'' of
varying intensities as illustrated in Fig.\,\ref{fig:diode_linearity_set_up}.
A 70/30 beam splitter following the stabilized light source extracts a constant
amount of light, the test signal, that is attenuated and guided through a
shutter into the integrating sphere. The direct light beam, the base signal,
passes through an aperture wheel allowing to vary base signal intensities. A
photodiode is mounted to the integrating sphere and, for different base
signals, its response to the base signal alone and to base signal plus test
signal is registered. The Si photodiode, which served as the the primary
monitoring photodiode for the reciprocity measurement, was used for this
calibration. It was illuminated\,\footnote{Pass-band selected light of
950$\,\pm\,$25\,nm was used.} at different intensities spanning five orders of
magnitude, and a power law model was fitted to evaluate the photodiode
linearity. In order to cover five orders of magnitude in  illumination, three
test signals of approximately 9\,pA, 55\,pA and 488\,pA were used as shown in
Fig.\,\ref{fig:diode_linearity}. The magnitudes of these test signals were
fitted along with a power law exponent, resulting in a non-linearity of
$(0.08\pm\,0.08)$\%/decade. This non-linearity was later utilized to correct
the detector response measurements, and its error was assigned as a systematic
uncertainty.

\begin{figure}[h]
\includegraphics[width=0.98\linewidth]{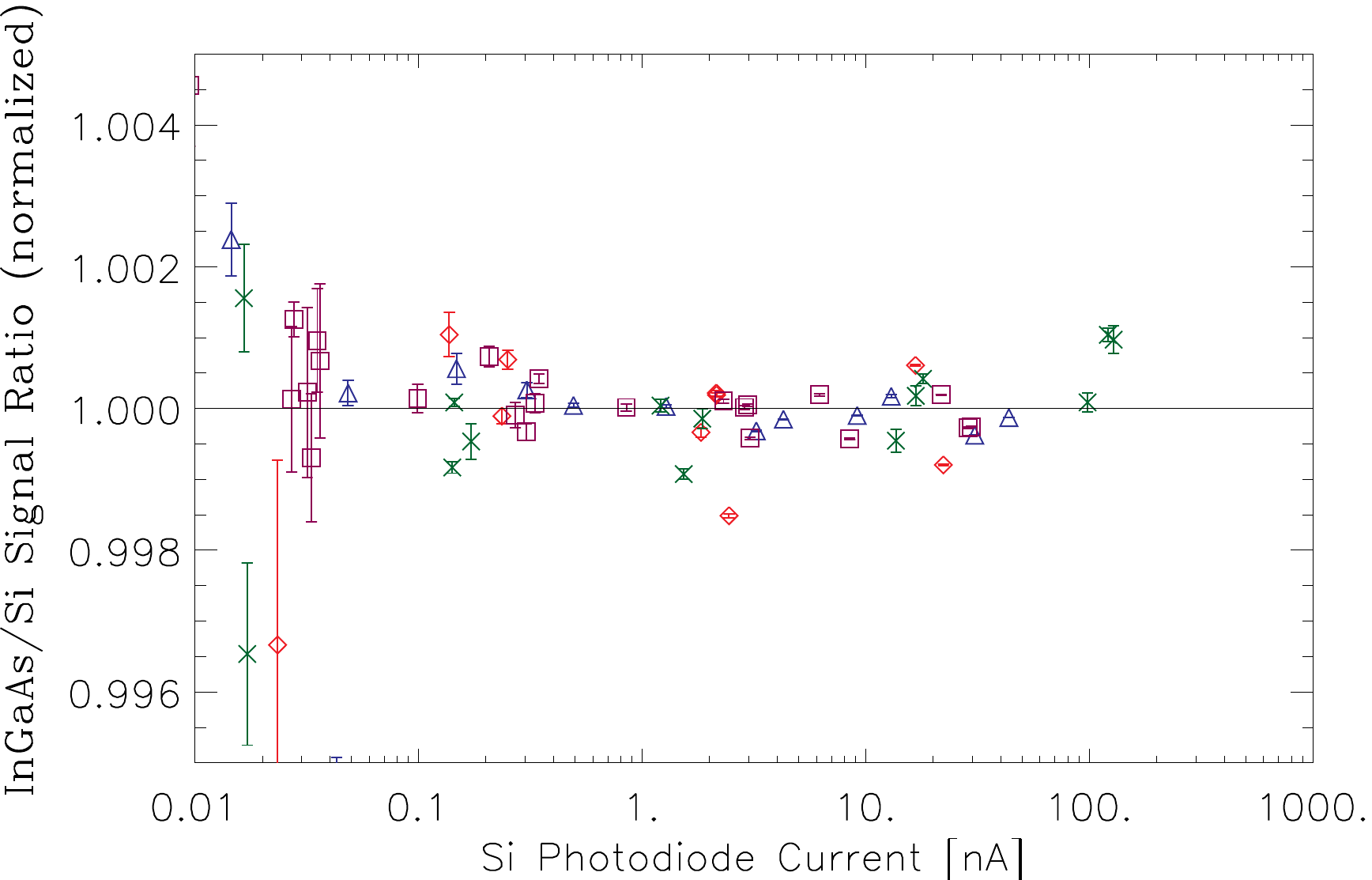}
\caption[diode_comparison]
{\label{fig:diode_comparison}
Normalized InGaAs photodiode to Si photodiode signal ratio as a function of Si photodiode current. Measurements at wavelengths of 700\,nm (red diamonds), 790\,nm (green crosses),
880\,nm (purple squares) and 950\,nm (blue triangles) are shown.  }
\end{figure}

As shown in Fig.\,\ref{fig:diode_comparison}, the \textit{relative} linearity
of the Si and InGaAs photodiodes is better than 0.1\% over the dynamic range of
illumination\,\footnote{The dynamic range corresponds to photodiode currents
between approximately 1\,pA and 100\,nA.} and wavelength used during the
reciprocity measurements. This agreement gives us confidence that the absolute
linearity of the InGaAs photodiode is also of the order of 0.1\%/decade, which
is consistent with previous linearity studies of Si and InGaAs photodiodes
\citep{Budde, Yoon}.

\subsection{Cryogenic System}

Reciprocity failure in NIR devices was characterized at a baseline temperature
of 140\,K in an 8-inch dewar manufactured by IR Labs. The hold time of the
system is typically 6 to 8\,hours, longer than the longest sampling sequence
which takes about 5\,hours to complete. This guarantees that measurements are
not disrupted by the liquid nitrogen refill process. For all measurements, the
NIR detector was mounted to a fixed copper heater plate which is weakly
thermally coupled to the liquid nitrogen reservoir and thermally stabilized to
10\,mK. The cool-down and warm-up ramp of 1\,K/min as well as temperature
stabilization of the NIR detectors at the operating temperature was controlled
and monitored by a precision temperature controller (Lakeshore 330). With the
temperature of the detector held constant at $140\,$K, the illumination system
inside the dewar was allowed to cool down to below $200\,$K at the integrating
sphere over a time period of about 8 hours. This is much colder than required
to suppress thermal background radiation in the $1.7\,\mu$m detector material.
A second temperature control loop was used to eliminate temperature dependence
in the response of the two photodiodes, which were always temperature
stabilized at $270\,$K.\footnote{It was observed that at lower temperatures the
InGaAs photodiode response becomes slightly non-linear.}

\subsection{Read-out and Control Electronics}

For detector read-out and control, a commercially available data acquisition
system from Astronomical Research Cameras (ARC) was used. In this system, 32
channels of parallel read-out are available from four 8-channel infrared video
processor boards combined with clock driver boards and a 250\,MHz timing and
PCI card. This read-out electronics is described in detail in \cite{Leach}.
Data are stored in FITS format for subsequent analysis. In the current set-up
no shutter was employed and thus each detector pixel starts to integrate signal
immediately after reset. Consequently, the shortest ``illumination time'' is
determined by the amount of time it takes to read the array. In the default
clocking mode (100\,kHz) the read-out of the whole array takes 1.418 seconds.
To reduce the illumination time, only a partial strip of the detector,
300\,$\times$\,2048 pixels was read out for most of the measurements. This
decreased the read-out time to 211 milliseconds.

Several detector characteristics depend on the bias voltage settings; the full
integration capacity for instance is a function of the reset voltage. All
measurements reported here were performed with bias settings that were
established to optimize low noise performance. The following voltages were
applied: detector substrate voltage Dsub\,=\,0.35\,V, reset voltage
Vreset\,=\,0.10\,V, pixel source follower bias voltage Vbiasgate\,=\,2.45\,V,
and pixel source follower source voltage Vbiaspower\,=\,3.23\,V.

\section{Measurements}

Following the discovery of dark energy in 1998, one of the earliest experiments
put forward to investigate the mysterious new property of the Universe was the
Supernova Acceleration Probe (SNAP),  a space-based telescope with a large wide
field imager comprised of CCD and NIR detectors. The project pursued a strong
detector procurement and development program for 1.7\,$\mu$m HgCdTe Focal Plane
Arrays (FPAs) with the goal of producing a low read-noise, high quantum
efficiency (QE) device suitable for the proposed instrument
\citep{schubnell2006}. Several lots were produced by TIS, with fabrication
based on the WFC3 development of 1\,k\,$\times$\,1\,k HgCdTe material. Low
read-noise and dark current, high QE as well as substrate removal were
addressed during different material growth runs.

One of the devices procured from TIS, H2RG-102, an engineering grade
2\,k\,$\times$\,2\,k 1.7\,$\mu$m cut-off detector mounted on a molybdenum
pedestal was characterized with our reciprocity set-up. This FPA is a
hybridized detector consisting of a highly integrated CMOS multiplexer and a
layer of infrared sensitive detector material. Photon conversion takes place in
a very thin layer of  HgCdTe, about 5 to 10\,$\mu$m thick (typically grown on a
much thicker substrate layer) with metallized contact pads defining the active
area. The accumulation of photogenerated electron-hole pairs on the junction
capacitance of the pixel causes a decrease in reverse bias which is sensed by a
MOSFET source follower. The multiplexer is an array of discrete read-out
transistors and, unlike a conventional CCD, can be read non-destructively.
Detector layer and multiplexer are indium bump-bonded and mounted to a pedestal
which equilibrates the temperature across the FPA.

\subsection{System Optimization}

Many challenges had to be overcome to achieve the 0.1\%/decade sensitivity to
reciprocity failure in our system. Initial testing of the set-up indicated that
it suffered from light leaks. The cryogenic ports identified as the source of
the leaks were shielded, and the internal baffling system was extended to fully
cover the detector to eliminate stray light in the system. The drifts in the
photodiode readout affecting low illumination measurements were first reduced
with better cable shielding and grounding, and finally corrected for in the
analysis using the signal from a reference photodiode. It was noticed that dark
images (where the aperture was closed) were brighter when the lamp was on than
when it was off. This was caused by the light heating the aperture mounts
causing them to glow in the NIR. It was mitigated by facing the reflective side
of the mounts towards the light and by using a cold short-pass filter between
the apertures and the integrating sphere for measurements below 1000\,nm. At
longer wavelengths, matched dark images were taken with the lamp on to allow a
complete subtraction of this small dark glow. One of the greatest challenges
involved the spectral mismatch of the detector and photodiode responses. The
comparison of the signals from both, the Si and InGaAs photodiodes indicated
that the pass-band filters leaked in the red. This was confirmed using a single
wavelength laser. Either short-pass or long-pass filters were placed in the
light path to improve out-of-band rejection. Monitoring photodiode signal
ratios also confirmed that the neutral density filters used were spectrally
flat to better than 0.1\% in the region we operated. This was not the case for
other neutral density filters we checked. Using apertures instead of neutral
density filters to control illumination avoids the spectral dependence issue.
Hence they were used as the primary means of illumination control. It turned
out, however, that the integrating sphere used was not large enough to fully
wash out the image of the aperture at its entrance and therefore different
apertures resulted in slightly different illumination patterns on the device.
This was remedied by two layers of spectrally flat diffusers, added between the
apertures and the integrating sphere. Ultimately, the different but
complementary means of attenuating the illumination, the apertures and the ND
filters, and the different spectral bands probed by the two photodiodes were
essential in reaching the required sensitivity in our measurements.

\subsection{Test and Analysis Procedure}

A comprehensive reciprocity test was performed on device H2RG-102. For all
measurements the detector  temperature was held constant at 140\,K.
Measurements were made with the QTH illumination system at wavelengths of
700\,nm, 880\,nm, 950\,nm and 1400\,nm and with the laser at 790\,nm.

H2RG-102 is a well performing engineering grade device.  The detector is
substrate removed and has an anti-reflective coating. The QE is greater than
90\% from 0.9\,$\mu$m to 1.7\,$\mu$m and the detector response extends far into
the visible: at 0.45\,$\mu$m QE is about 40\%. The dark current and read-noise
performance is good, with a Fowler-16 noise of $10\,e^-$ for a $300\,$s
exposure at $140\,$K.

During a typical reciprocity measurement the detector was first reset then
repeatedly read non-destructively in a procedure that is generally called
``Sample Up the Ramp'' (SUR) mode, with up to 200 frames read during an
exposure. For every SUR sequence ``matched darks'' were obtained. Measurement
conditions for the matched darks were in every way identical to the reciprocity
measurement conditions but exposures were taken with the aperture closed. The
integrated signal, $S$, in the detector is parameterized as
$S(t,F)=\int_0^{t}{F(t^\prime)\times\epsilon(S){\rm d}t^\prime}$, where $F(t)$
represents the detector count-rate as a function of time $t$, and $\epsilon(S)$
takes into account classical non-linearity (for details see Appendix). As can
be seen in Fig.\,\ref{fig:total_signal}, applying this parametrization
describes the observed behavior well. After the correction, the classical
non-linearity for signals below 60\% of the saturation level is less than
0.1\%. Exposed images and matched darks were then included in the fit procedure
used in calculating the NIR detector response.
\begin{figure}[h]
\begin{center}
\begin{minipage}[b]{0.98\linewidth}
\includegraphics[width=0.97\linewidth]{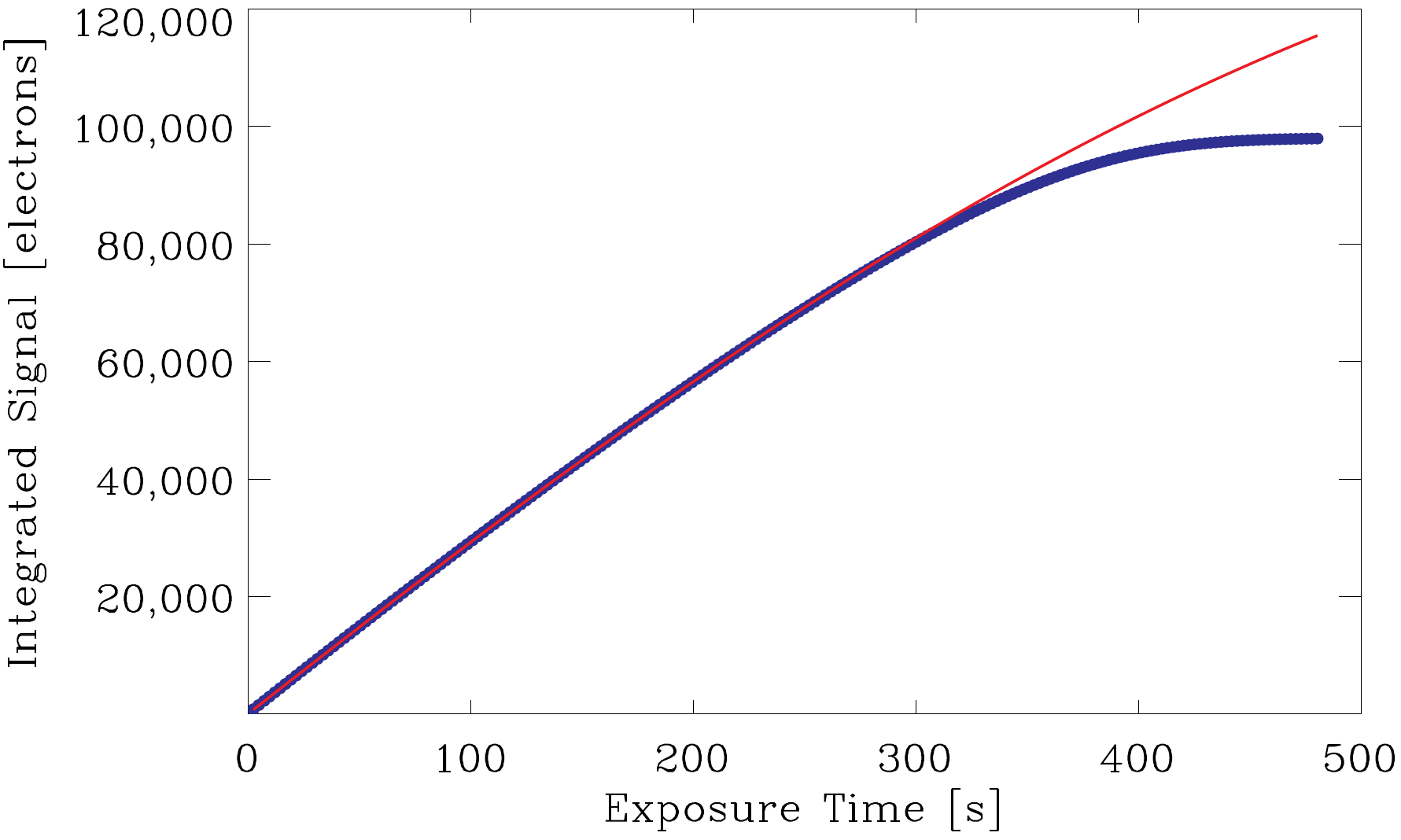}
\vspace*{2mm}
\end{minipage}
\hspace*{4.5mm}
\begin{minipage}[b]{0.98\linewidth}
\includegraphics[width=0.95\linewidth]{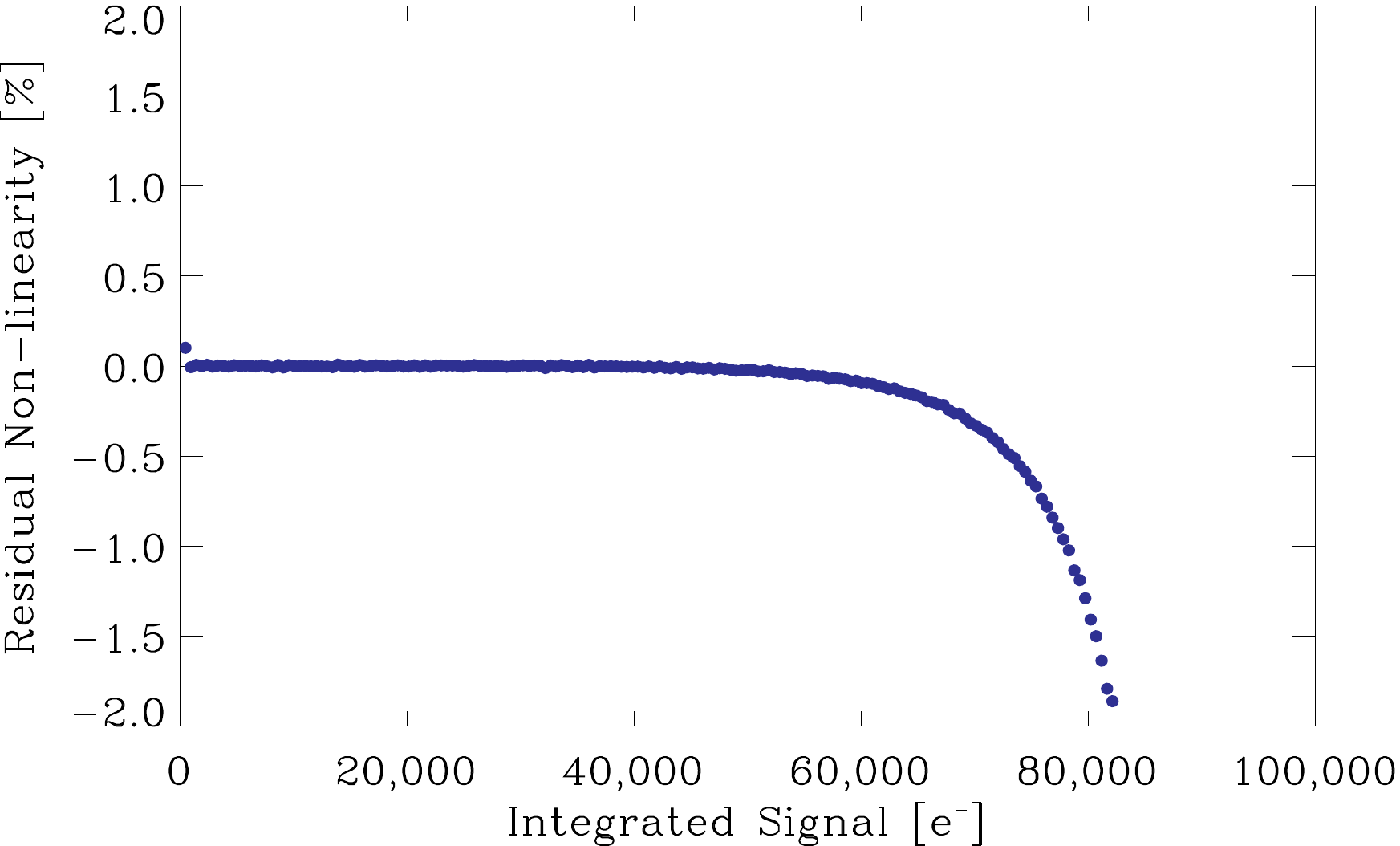}
\end{minipage}

\caption[total_signal]{
Upper panel: Integrated signal in a HgCdTe detector as a function of time.
The red curve is the result of the three-parameter fit described in the Appendix.
Lower panel: Deviation of the data from the fit versus integrated charge.
The residual non-linearity is reduced to below 0.1\% (1\%) for signals below
60\% (80\%) of the saturation level.} \label{fig:total_signal}
\end{center}
\end{figure}

\begin{figure}[h]
\begin{center}
\begin{minipage}[b]{0.98\linewidth}
\begin{center}
\begin{tabular}{c}
\includegraphics[width=\linewidth]{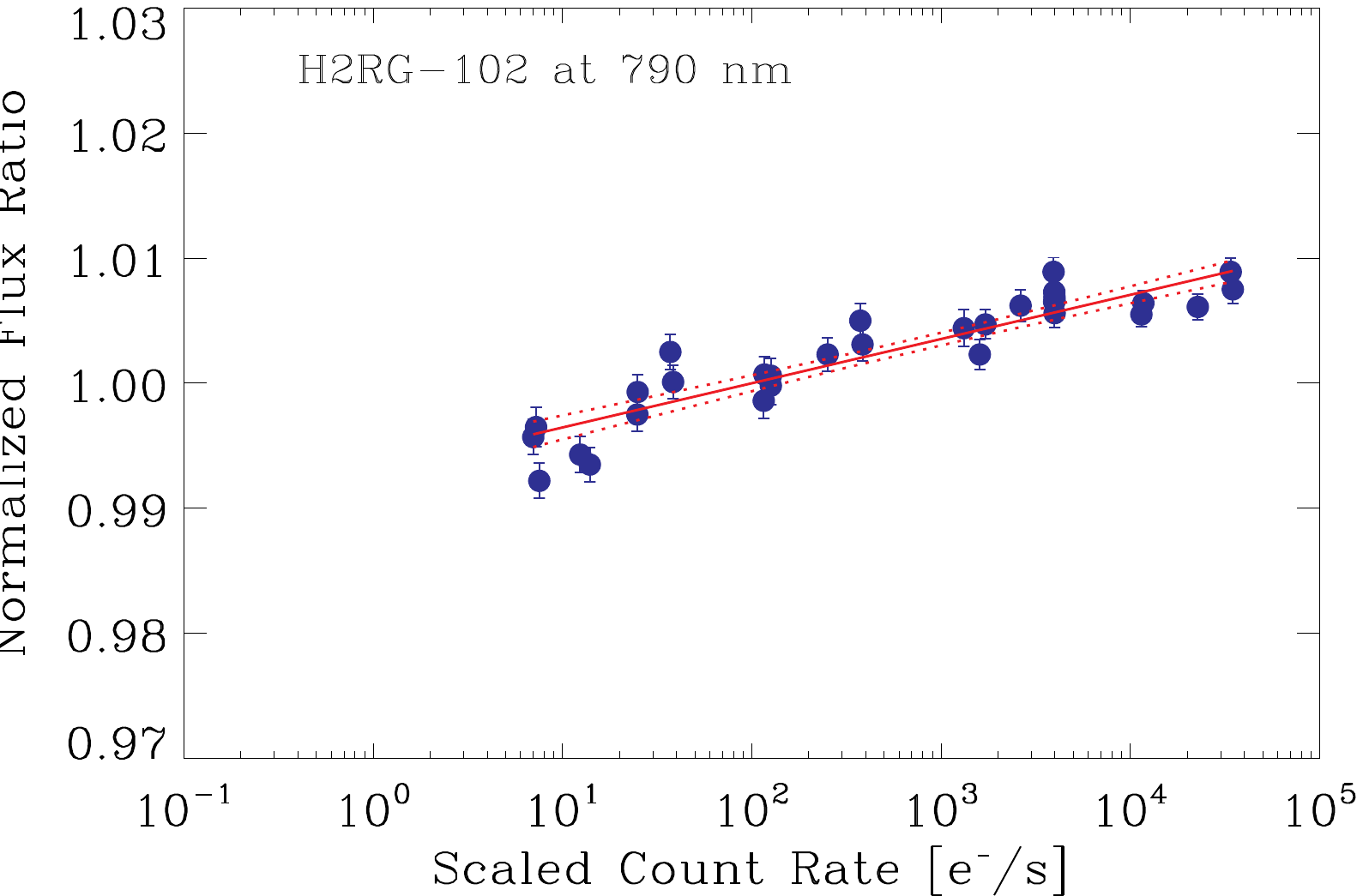}
\end{tabular}
\end{center}
\vspace*{2mm}
\end{minipage}
\begin{minipage}[b]{0.98\linewidth}
\begin{center}
\begin{tabular}{c}
\includegraphics[width=\linewidth]{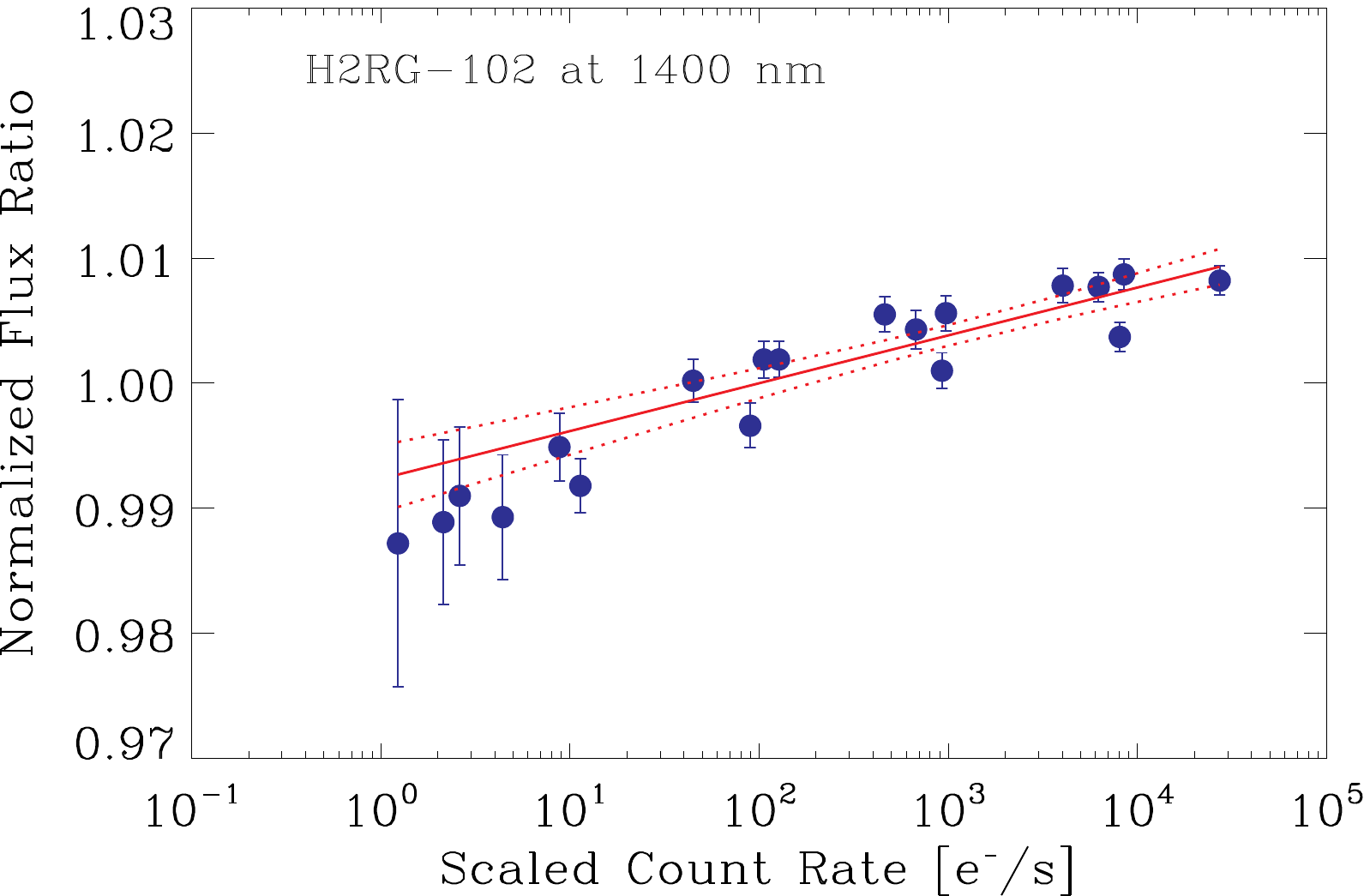}
\end{tabular}
\end{center}
\caption[rec_failure_236]{Reciprocity failure versus scaled count rate in device H2RG-102
at 790\,nm (upper panel) and 1400\,nm (lower panel).  The solid lines indicate a logarithmic
fit to the data points. The 1$\sigma$ error bands (dotted lines) include the point-to-point statistical and systematic uncertainties, but not the systematic uncertainty due the photodiode calibration of 0.08\%/decade.  The measured values for the reciprocity failure at 790\,nm is
($0.35\,\pm\,0.03\,(\mbox{stat.})\,\pm\,0.08\, (\mbox{syst.})$)\%/decade, and ($0.38\,\pm\,0.05\,(\mbox{stat.})\,\pm\,0.08\, (\mbox{syst.})$)\%/decade at 1400\,nm.}
\label{fig:rec_failure_102}
\end{minipage}
\end{center}
\end{figure}

Monitoring photodiode currents were recorded for each frame in the sample and
corrected as shown in Fig.\,\ref{fig:photodiode}. Long exposures over several
hours were typical at the lowest illumination levels of a few
electrons/pixel/second at the detector. It was observed that at the most
sensitive setting the pico-ammeter drifts at the 10\% level. Those fluctuations
were tracked by a reference photodiode connected to a pico-ammeter and removed
from the data. The residual variation in the current measurement is dominated
by statistical fluctuations and the variance of the mean improves linearly with
the number of measurements in the exposure. The Si photodiode itself was found
to deviate from linearity at a level of $(0.08\pm0.08)$\%/decade, requiring a
correction that reduced the photodiode signal by this amount. The uncertainty
in the Si photodiode calibration along with the InGaAs to Si photodiode ratios
constitute the systematic limit of our sensitivity to reciprocity failure of
0.1\%/decade. The fitted detector response is divided by the photodiode current
resulting in the flux ratios shown in Fig.\,\ref{fig:rec_failure_102}.
Normalized flux ratios were obtained at different illumination intensities and
at different wavelengths. At wavelengths below 1000\,nm, current readings from
the Si photodiode and above 1000\,nm, readings from the InGaAs photodiode were
used for calculating the flux ratios.

In our ad-hoc model, the two non-linearity parameters were fitted
simultaneously to all the different illumination intensity sets, while the flux
was fit separately. This ensures that reciprocity failure is not hidden in the
possible degeneracy of those parameters. It also reduces the uncertainties on
the estimated parameters. As a check we also fitted each illumination set
separately. The values for reciprocity failure so obtained agreed with the
combined fit results. Details of the parametrization are discussed in the
Appendix.

\subsection{Results}

\begin{figure}[h]
\begin{center}
\begin{minipage}[b]{0.98\linewidth}
\begin{center}
\begin{tabular}{c}
\includegraphics[width=\linewidth]{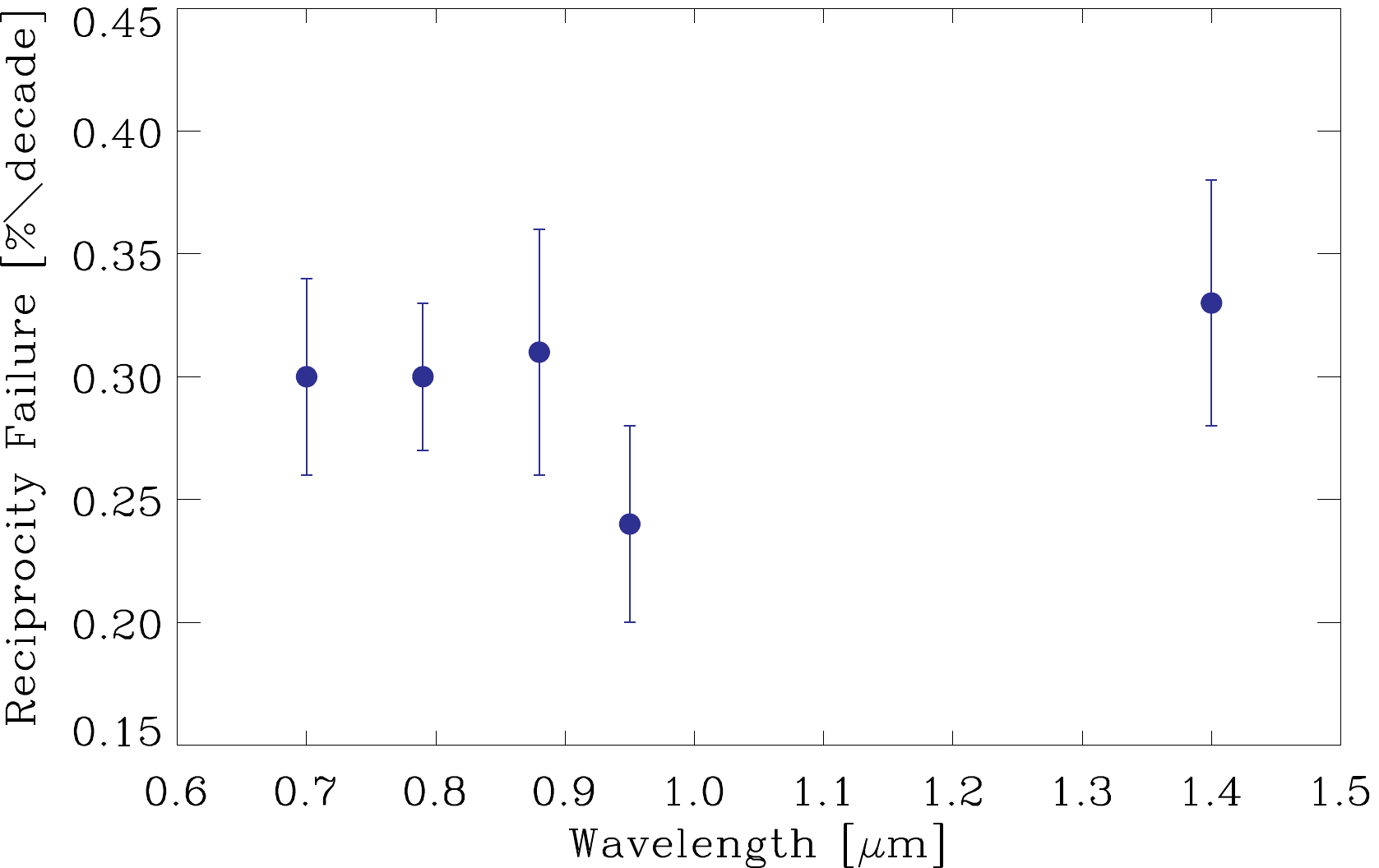}
\end{tabular}
\end{center}
\end{minipage}
\caption[rec_failure_wave]{Reciprocity failure as a function of wavelength for device H2RG-102.}
\label{fig:rec_failure_wave}
\end{center}
\end{figure}

Figure\,\ref{fig:rec_failure_102} shows the flux ratios as a function of count
rate with a logarithmic fit (linear in log illumination) that describes the
data well. As indicated in the figure, reciprocity failure for the H2RG-102
detector tested in our set-up is very low. Measurements were performed at five
different wavelengths (700\,nm, 790\,nm, 880\,nm, 950\,nm and 1400\,nm) with no
significant wavelength dependence observed as shown in
Fig.\,\ref{fig:rec_failure_wave}. The NIR detector count rate is scaled
relative to the photodiode current to remove flux dependence from the
horizontal axis. Measured values for the reciprocity failure at the five
wavelengths (in \%/decade) are 0.35$\,\pm\,$0.04, 0.35$\,\pm\,$0.03,
0.36$\,\pm\,$0.04, 0.29$\,\pm\,$0.04, and 0.38$\,\pm\,$0.05. These reciprocity
failure values are subject to a 0.08\%/decade systematic uncertainty in the
photodiode non-linearity correction. This result contrasts with the strong
wavelength dependence for reciprocity failure in all three NICMOS detectors.

\section{Summary}

We have built a test station for the measurement of reciprocity failure in NIR
detectors and achieved a sensitivity of approximately 0.1\%/decade. Initial
measurements were performed on a 1.7$\,\mu$m HgCdTe detector (HR2G-102) between
700\,nm and 1400\,nm which yielded a non-linearity due to reciprocity failure
of about 0.35\%/decade. We find no indication for wavelength dependence in the
tested detector. This contrasts with the reported behavior of the NICMOS
detectors on HST. The fabrication of JDEM/SNAP devices is based on WFC3
detector development. This is reflected in measurements on the final candidate
detectors for WFC3 which show very similar results as H2RG-102
\citep{Hill_Garching}. The WFC3 team reports reciprocity failure ranging from
0.3\%/decade to 0.97\%/decade for three detectors. As they point out, this is
significantly smaller than the effect seen for the 2.5\,$\mu$m HgCdTe NICMOS
detectors on HST (6\%/decade). NICMOS detector material was grown using the
liquid phase epitaxy technique while molecular beam epitaxy was used for the
growth of material for the WFC3 and for the JDEM/SNAP R\&D detectors.

We plan to extend these measurements to a variety of detectors and to use
spatial maps and temperature dependence of reciprocity failure to investigate
this effect further.

\acknowledgments
We gratefully acknowledge the many valuable conversations with
Roger Smith and Christopher Bebek during the course of this work. We also thank
the reviewer for many insightful and constructive comments. This work was
supported by the Director, Office of Science, of the U.S. Department of Energy
under Contract Nos. DE-FG02-95ER40899 and DE-FG02-08ER41566.


\section{Appendix}

In order to properly evaluate detector response at differing illumination
intensities, care must be taken to distinguish between reciprocity failure and
classical non-linearity as the pixel integrates charge.

An ad-hoc three parameter model, intrinsically independent of the intensity
level, was produced to describe the change in junction capacitance of the pixel
as a function of integrated  signal $S$. In a perfectly linear detector the
voltage changes by a constant amount for each collected electron until the
voltage is sufficient to forward-bias the detector diode. In a real detector
this voltage change decreases with increasing $S$. Two parameters, $a$ and $b$
are used to parameterize this behavior, such that
\begin{align}
\epsilon(S) = \frac{a + 1 - (a + 1)^{\frac{S}{b}}}{a}\,, \label{E:cce}
\end{align}
where $\epsilon$ is defined to be unity when no charge, $S$, has been collected
$(S = 0)$,  and zero when the pixel has ``saturated'' $(S = b)$. The parameter
$a$ describes how quickly the junction capacitance is changing, as $a \to
\infty$ the device becomes linear. The parameter $b$ is the maximum voltage
that the pixel can record, that is, the pixel saturation level. The rate of
signal integration by the device can be written as
\begin{align}
\frac{dS}{dt} = F(t) \epsilon(S)\,, \label{E:dsdt}
\end{align}
where $F(t)$ is the time dependent true flux.

Equation (\ref{E:dsdt}) can be integrated analyticaly only for certain models
of the  flux $F(t)$. We approximate the flux as constant illumination plus a
dark current (with constant asymptotic value, $d$, and an exponentially
decaying component, $d_{e}$). The flux can then be written as
\begin{align}
F(t) = F_0 + d + d_{e}\cdot exp({-\frac{t}{\tau}})\,.  \label{E:flux}
\end{align}
The dark current is fitted separately with the exponentially decaying model
using  data sets obtained in the dark resulting in the values of $d$, $d_{e}$,
and $\tau$ being known at the time of the integrated signal fit.

Equation (\ref{E:dsdt}) is then integrated to the form
\begin{align}
S(t) = \frac{b}{\log{(1 + a)}}
\log{\left(\frac{1 + a}
{1+\exp{\left(\frac{\alpha}{b} + \frac{\alpha}{ab} + \frac{\beta}{b}\right)}}\right)}\,, \label{E:integ}
\end{align}
with $\alpha$ and $\beta$ defined as
\begin{equation}
\alpha = \left(d + F_0 - dt - F_0t + d_{e}\tau(e^{-\frac{t}{\tau}} - e^{-\frac{1}{\tau}})\right)
\log{(1 + a)} \,, \label{E:alpha}
\end{equation}
\begin{eqnarray}
\beta &=& \left(\left(d + F_0 + d_{e}\tau(1 - e^{-\frac{1}{\tau}})\right)
(-1 - \frac{1}{a}) + \frac{b\log{a}}{\log{(1 + a)}}\right) \nonumber \\ &&\cdot\log{(1 + a)}\,.  \label{E:beta}
\end{eqnarray}
After discarding the first frame to avoid turn-on effects, each $i^{th}$ SUR
image, $S(t_{i}) - S(t_{i-1})$, is fitted for the three parameters, $a, b$ and
$F_0$.  The value of $F_0$ serves as the detector response independent of the
classical non-linearity and is divided by the corresponding photodiode current
to compute the normalized flux ratio.

\end{document}